\documentclass{article}
\usepackage{amsmath}
\usepackage{amssymb}
\usepackage{color}
\usepackage{authblk}

\usepackage{vmargin}                   
\setmarginsrb{1.5cm}{1.5cm}{1.5cm}{1.5cm}{0cm}{1cm}{0cm}{1cm}   

\usepackage{graphicx}
\usepackage{bm}

\newcommand\rr{{\bf r}}
\newcommand\ee{{\bf e}}
\author[1]{Vincent Tejedor}
\affil[1]{Cour des comptes, 13, rue Cambon, 75 001 Paris, France}

\begin{document}
\title{A dimensional acceleration of gradient descent-like methods, using persistent random walkers}

\maketitle

\section*{Introduction}

Finding a local minimum or maximum of a function is often achieved through the gradient-descent optimization method\cite{Curry:1944}. If the function $f$ to be optimized is differentiable, a way to quickly find its (local) minimum is to follow the steepest slope, given by the opposite of the gradient direction. Some variants can be used to accelerate the convergence algorithm\cite{Fletcher:1963}. For a function of $d$ variables, the gradient requires to compute at each step $d$ partial derivatives of the $f$ function. When the function $f$ is not known analytically, each partial derivative is approximated by a local slope obtained with two values of $f$ at $x_i-dx_i$ and $x_i+dx_i$\cite{Cauchy:1847}.

This method is for instance used in machine-learning, to fit the models parameters so as to minimize the error rate on a given data set. Since each step requires to obtain $d$ partial derivatives of the $f$ function, it can become time-consuming when $d$ grows and when each computation of the $f$ function is complex. A variant of the gradient-descent method has been developped, when $f$ is the sum of differentiable functions $f_i$, by computing the gradient only on a limited subset of $f_i$. By switching the subset at each step, the stochastic gradient method allows to limit the computation time at each step\cite{Bottou:2008}, while ensuring a convergence toward the local minimum.

However, if the computation time of $f$ is the limiting factor, the convergence process can still be optimized using persistent random walks. For all the gradient-related method, we here propose a way to minimize the optimization process by using random walks instead of gradient computing. Optimization works here on the dimensional aspect of the function and not on the set size: this approach can thus be combined with algorithm improvement based on the set size. As shown in a previous publication, the random walk can be further optimized with persistence\cite{Tejedor:2012}.

We here detail the method principle, show an estimate of the acceleration factor and check numerically that this estimation is valid for quadratic functions.

\section{Model}

The principle of the persistent random descent is to choose at each step a random direction among the vectors of an orthonormal basis $(\ee_1, \ldots, \ee_d)$, to check if this direction allow to lower $f$ value, and if so to continue in this direction for a random number of steps as long as $f$ disminushes, before choosing once again a random direction. The walk is random in the sense that the direction is chosen randomly, and persistent since once a direction is choosen, this direction is privileged at the next steps.

The algorithm can be described as follow:
\begin{enumerate}
\item Position is  $\rr$. Choose a random direction $\delta^{i=1}_{\rm step} \in \{ \pm \ee_1, \ldots, \pm \ee_d \}$.
\item If $f(\rr + \delta^i_{\rm step}) < f(\rr)$ continue, else choose another direction $\delta^{i+1}_{\rm step} \in \{ \pm \ee_1, \ldots, \pm \ee_d \} \setminus \{ \delta^k_{\rm step} \}_{k=1 \ldots i}$ and redo this step. If $i= 2 d$, exit the algorithm and return $\rr$ as a (presumed) local minimum.
\item Generate a random integer $t_{\rm pers}$, and for $j \in [1, t_{\rm pers}-1 ]$, as long as $f(\rr + (j+1) \delta^i_{\rm step}) < f(\rr + j \delta^i_{\rm step})$, continue. Else stop, set $\rr =  \rr + j \delta^i_{\rm step}$ and go back to step 1.
\end{enumerate}

Basically, at each step, different directions among the one of an orthonormal basis are tested. The first one leading to a descent is kept, and the walk persists in this direction for a random time $t_{\rm pers}$. As soon as this direction is not descending or as the number of steps exceeds $t_{\rm pers}$, a new sensing phase among the direction of the orthonormal basis is launched to choose the new first descending direction. The algorithm ends as soon as it finds a point where no step along a direction of the orthonormal basis allow to descend anymore.

\section{Speed of convergence}

To compare this method with a classical gradient-descent model, one assume that the computation of $f$ is the limiting factor in computing time. 

Each step for a gradient-descent model consists in computing $2 d$ values of $f$, namely $f(\rr \pm \ee_i)$ with $ i \in [1, d]$, in order to estimate the gradient by differenciation. The size of the step is then assumed to be $1$ (this assumption can be modified in both methods, so that the relative difference is not changed). These two values are modified in the persistent random descent model: less $f$ computations are needed, but the effective step length, defined as the distance variation to the targeted optimum at each step, is less than 1, since the direction chosen is not the steepest one.

\subsection{Locally quadratic assumption}
The speed of convergence is first determined with a locally quadratic assumption, or less restrictively when at each step, half of the space is going downward and the other half upward.

\subsubsection{Number of computation of $f$ at each step}

For the persistent random descent, each step consists in two phases. First, one try successively all directions of an orthonormal basis $(\ee_1, \ldots ,\ee_d)$. If one assume that $f$ is locally quadratic: half of the space is going downward, the other half upward. Each direction has then a probability $1/2$ to be selected: one has to find one of the $d$ directions goeing downward in a sample of size $2d$. With replacement, the average number of try to find one of the $d$ good direction is given by:
\begin{equation}
\langle {\rm step} \rangle_{\rm with \ replacement} = \sum_{i = 1}^{\infty} i \left ( 1 - \frac{1}{2} \right )^{i-1} \frac{1}{2} = \sum_{i = 1}^{\infty} \frac{i}{2^i} = 2
\end{equation}
The algorithm being without replacement, this average number of steps can be lowered:
\begin{equation}
\langle {\rm step} \rangle = \sum_{i = 0}^{d} (i+1) \left ( \frac{d}{2d-i} \prod_{j=0}^{i-1} \left ( 1 - \frac{d}{2d - j} \right )  \right ) \underset{n \to \infty}{\to} 2
\end{equation}
Convergence for large $d$ can be obtained by noting that:
\begin{align}
\prod_{j=0}^{i-1} \left ( 1 - \frac{d}{2d - j} \right ) & = \frac{1}{2^i} \prod_{j=0}^{i-1} \frac{d-j}{d - \frac{j}{2}}\\
\ln \left (  \prod_{j=0}^{i-1} \frac{d-j}{d - \frac{j}{2}} \right ) & =  \sum_{j=0}^{i-1} \ln \left ( \frac{d-j}{d - \frac{j}{2}} \right )\\
& \underset{d \to \infty}{\to} \sum_{j=0}^{i-1} \left ( - \frac{j}{2 d } + \mathcal{O} \left ( \left ( \frac{j}{d} \right )^2 \right ) \right )\\
& \underset{d \to \infty}{\to}  - \frac{i^2}{4 d} + \mathcal{O} \left (  \frac{i^3}{d^2} \right ) 
\end{align}
This approximation is only valid when $i \ll d$, which will be the case for the useful part of the overall sum.
\begin{equation}
\langle {\rm step} \rangle = \sum_{i = 0}^{d} \frac{i+1}{2^{i+1}} \left ( \frac{1}{\displaystyle 1-\frac{i}{2 d}} \prod_{j=0}^{i-1} \frac{d-j}{d - \frac{j}{2}}  \right ) \underset{d \to \infty}{\to} \sum_{i = 0}^{d} \frac{i+1}{2^{i+1}} \left ( \frac{\displaystyle \exp \left ( - \frac{i^2}{4d} \right ) }{\displaystyle 1-\frac{i}{2 d}}  \right )
\end{equation}
The last part of the sum, in parenthesis, is close to one for $i \ll d$, and goes to $0$ for large $i$. Since the first part of the sum decreases like $2^i$, one can have an estimate of the average number of steps as follow:
\begin{equation}
\langle {\rm step} \rangle  \underset{d \to \infty}{\to} \sum_{i = 0}^{d} \frac{i+1}{2^{i+1}} \left ( 1  - \frac{i^2}{4d} + \frac{i}{2 d} \right ) = 2 \left ( 1 - \frac{1}{d} \right ) + \mathcal{O} \left ( \frac{1}{d^2} \right )
\end{equation}
As the number of dimension $d$ increases, the average number of try to find a decreasing direction, under a locally quadratic assumption, rises to $2$ (value obtained with replacement). This average has to be compared to the $2 d$ of the gradient descent approach.

\subsubsection{Efficiency of the direction chosen}

The gradient descent always choose the steepest slope, whereas the persistent random model choose the first direction goeing downward. This direction is in average not the steepest one. Once again, under the locally quadratic assumption, one can estimate that the efficiency of each approach is related to its projection along the steepest slope. The gradient descent method is 100~\% efficient (by definition), the persistent random model choose any direction of the orthonormal basis goeing downward. To compute the projection of the random direction chosen on the steepest slope, one use hyperspherical coordinates with $r = 1$ (orthonormal basis):
\begin{equation}
\left ( \begin{array}{l} x_1 \\ x_2 \\ \vdots \\ x_{d-1} \\ x_d \end{array}
\right ) = \left ( \begin{array}{l} \cos \left ( \theta_1 \right ) \\ \sin \left ( \theta_1 \right ) \cos \left ( \theta_2 \right ) \\ \vdots \\ \sin \left ( \theta_1 \right ) \ldots \sin \left ( \theta_{d-2} \right ) \cos \left ( \theta_{d-1} \right )  \\ \sin \left ( \theta_1 \right ) \ldots \sin \left ( \theta_{d-2} \right ) \sin \left ( \theta_{d-1} \right ) \end{array} \right )
\end{equation}
, where $(\theta_1, \ldots , \theta_{d-1}) \in [0, \pi]^{d-2}$ and $\theta_{d-1} \in [0, 2 \pi]$. Under the locally quadratic assumption, the steepest slope in anywhere in the half space goeing downward. One can thus consider that the projected length along this steepest direction is:
\begin{equation}
\langle {\rm length} \rangle = \frac{\displaystyle \left ( \int_{0}^{\pi} \right )^{d-1} \rr . \left (\begin{array}{l} 0 \\ 0 \\ \vdots \\ 0 \\ 1 \end{array} \right )  dS}{\displaystyle \left ( \int_{0}^{\pi} \right )^{d-1} dS} =  \frac{\displaystyle \left ( \int_{0}^{\pi} \right )^{d-1} \sin \left ( \theta_1 \right )^d \ldots \sin \left ( \theta_{d-2} \right )^{2} \sin \left ( \theta_{d-1} \right )  d\theta_1 d\theta_2 \ldots d\theta_{d-1}}{\displaystyle \left ( \int_{0}^{\pi} \right )^{d-1} \sin \left ( \theta_1 \right )^{d-1} \ldots \sin \left ( \theta_{d-2} \right )  d\theta_1 d\theta_2 \ldots d\theta_{d-1}}
\end{equation}
The upper term of the fraction is the product between the $(0,0,\ldots , 0, 1)$ vector (steepest slope direction, in an adequate orthonormal basis) and a random vector in hyperspherical coordinates that describes half a space around this vector. The lower term is the surface of this half space at $r=1$. This expression can be simplified, and behavior for large $d$ obtained using Stirling' approximation:
\begin{equation}
\langle {\rm length} \rangle = \frac{\displaystyle \int_{0}^{\pi} \sin \left ( \theta_1 \right )^d  d\theta_1}{\displaystyle \int_{0}^{\pi} d\theta_1} = \frac{1}{\sqrt{\pi}} \frac{\displaystyle \Gamma \left ( \frac{1 + d}{2} \right ) }{\displaystyle \Gamma \left (1 + \frac{d}{2} \right ) } \underset{d \to \infty}{\to} \sqrt{\frac{2}{\pi d} } \left ( 1 - \frac{1}{4 d} + \mathcal{O} \left ( \frac{1}{d^2} \right ) \right )
\end{equation}

\begin{figure}[htb!]
\centering \includegraphics[width = 0.5\linewidth,clip]{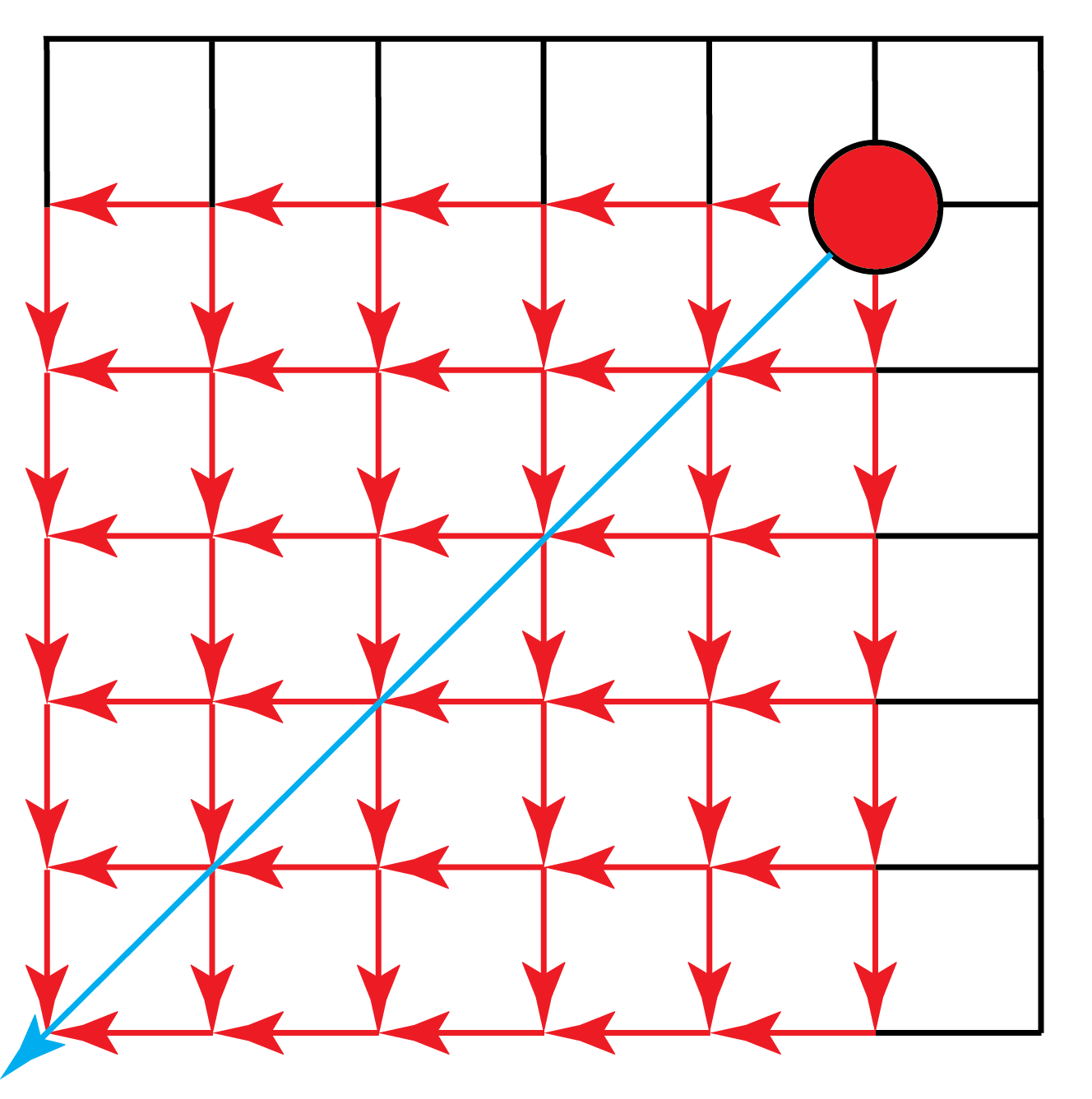} 
\caption{(color online) Illustration in $d = 2$ of the difference between trajectories of the gradient descent model (long blue arrow) and the persistent random walk model (lattice of short red arrows) for a $x^2 + y^2$-like problem (locally quadratic).}
\label{fig:dist}
\end{figure}

\subsubsection{Speed of convergence for locally quadratic function}

By combination of the two previous results, the comparative speed of the two research strategies, defined as the ratio of the length of each step divided by the computation time at each step, is:
\begin{equation}
{\rm acceleration}_{\rm locally \ quadratic} = \frac{\langle {\rm speed} \rangle_{PRW}}{\langle {\rm speed} \rangle_{\rm gradient}} = \frac{\displaystyle \frac{\langle {\rm length} \rangle}{\langle {\rm step} \rangle}}{\displaystyle \frac{1}{2 d}}  \underset{d \to \infty}{\to}   \sqrt{\frac{2}{\pi} d} \left ( 1 +  \frac{3}{4 d} + \mathcal{O} \left ( \frac{1}{d^2} \right ) \right )
\end{equation}

In order of magnitude, the acceleration factor increases as $\sqrt{d}$, where $d$ is the space dimension. For machine-learning problems, where $d$ is typically above 1000, the acceleration factor can be significative.

\subsection{Single possibility assumption}

On the other side of the spectrum, one can assume that only one direction goes downward, the $2 d -1$ others being rejected.

\subsubsection{Number of computation of $f$ at each step}

Under the single possibility assumption, the average number of try to find the correct direction can be obtained analytically:
\begin{equation}
\langle {\rm step} \rangle = \sum_{i = 0}^{2d - 1} (i+1) \left ( 1 - \frac{1}{2d - i} \right ) \prod_{j=0}^{i-1} \left ( 1 - \frac{1}{2 d -j} \right ) = \sum_{i = 0}^{2d - 1} \frac{i+1}{2d} =  d + \frac{1}{2}
\end{equation}

\subsubsection{Efficiency of the direction chosen}

To compare the gradient descent model and the random persistent model, one has to recall that the gradient is estimated using $2 d$ computations of $f$: if all direction but one are goeing upward, the estimated gradient should point to the direction goeing downward. Both model will thus have the same efficiency by leading to the same direction.

\subsubsection{Speed of convergence under single possibility assumption}

By combination of the two previous results, the comparative speed of the two research strategies, defined as the ratio of the length of each step divided by the computation time at each step, is, under the single possibility assumption:
\begin{equation}
{\rm acceleration}_{\rm single \ possibility} = \frac{\langle {\rm speed} \rangle_{PRW}}{\langle {\rm speed} \rangle_{\rm gradient}} = \frac{\displaystyle \frac{\langle {\rm length} \rangle}{\langle {\rm step} \rangle}}{\displaystyle \frac{1}{2 d}}  = 2 \frac{2 d}{2 d + 1} \underset{n \to \infty}{\to} 2
\end{equation}

In order of magnitude, the acceleration factor is here simply $2$, no matter the space dimension $n$. When only one solution exists, searching randomly increases slightly the speed of convergence, but slightly. 

One can however note that so far, the persistence effect has been neglected: for smooth functions, where several steps in the same direction can be made toward the local optimum, the number of computation at each step switch from $d + 1/2$ to $1$, leading to a $2 d$ acceleration factor. Even if this effect depends on the function topology, it can only accelerate (with a huge coefficient when $d$ is large) the speed of convergence.

\subsection{Generic case}

The generic assumption could be that $k \in [1,2d]$ directions are goeing downward, $k = d$ being the locally quadratic assumption, and $k=1$ being the single possibility assumption.

\subsubsection{Number of computation of $f$ at each step}

As previously, one can compute the average number of steps with the following formula:
\begin{equation}
\langle {\rm step} \rangle = \sum_{i = 0}^{2 d - k} (i+1) \left ( \frac{k}{2d-i} \prod_{j=0}^{i-1} \left ( 1 - \frac{k}{2d - j} \right )  \right )  \underset{d \to \infty}{\to} \frac{2 d}{k}
\end{equation}

\subsubsection{Efficiency of the direction chosen}

As a first step, one can find an upper bound to the length ratio between the two methods, by comparing the chemical distance $(\left \lVert \vec{\rr} \right \rVert_1)$ with the Euclidian distance $(\left \lVert \vec{\rr} \right \rVert_2)$. The gradient descent always choose the steepest slope while the persistent random model follows a direction of the orthonormal basis: the later one perform a random walk on an Euclidian lattice (defined by the orthonormal basis), where the chemical distance is the number of steps performed on the lattice, while the Euclidian distance is the classical distance on an Euclidian space of dimension $d$ (see Fig. \ref{fig:dist}). One can note that:
\begin{equation}
\left \lVert \vec{\rr} \right \rVert_1^2 = \left \lVert \left ( \begin{array}{l} x_1 \\ x_2 \\ \vdots \\ x_{d-1} \\ x_d \end{array} \right ) \right \rVert_1^2 = \left ( \sum_{i=1}^d \sqrt{x_i^2} \right )^2 = \sum_{i=1}^d \sum_{j=1}^d \left ( \sqrt{x_i^2} \sqrt{x_j^2} \right ) \le \sum_{i=1}^d \sum_{j=1}^d \frac{x_i^2 + x_j^2}{2} = d \sum_{i}^d x_i^2 = d \left \lVert \vec{\rr} \right \rVert_2^2
\end{equation}
Using a lattice instead of the continuous space thus lead to a distance that is at most $\sqrt{d}$ times greater: the relative efficiency of the persistent random model compared to the gradient method should thus be at least $1/\sqrt{d}$ on the length point of view.

If $k \in [1,d]$, one can use the previous result for locally quadratic function, where only $k$ direction are covered instead of $d$: other directions can not be chosen by the random walker or by the gradient descent model. The length is then:
\begin{equation}
\langle {\rm length} \rangle = \frac{\displaystyle \int_{0}^{\pi} \sin \left ( \theta_1 \right )^k  d\theta_1}{\displaystyle \int_{0}^{\pi} d\theta_1} = \frac{1}{\sqrt{\pi}} \frac{\displaystyle \Gamma \left ( \frac{1 + k}{2} \right ) }{\displaystyle \Gamma \left (1 + \frac{k}{2} \right ) } \underset{k \to \infty}{\to} \sqrt{\frac{2}{\pi k} }
\end{equation}
The factor $\sqrt{2/\pi}$ that remains even for $k = 1$ is related to the fact that the motion is performed on an Euclidian discrete lattice. The smaller length is reached for $k = d$, namely the locally quadratic case.

If $k \in [d, 2d]$, integration has to be performed for some dimensions on $[0, 2 \pi]$ instead of $[0, \pi]$, leading also to a dimensional restriction. The length is in such case:
\begin{equation}
\langle {\rm length} \rangle = \frac{\displaystyle \int_{0}^{\pi} \sin \left ( \theta_1 \right )^{2d - k}  d\theta_1}{\displaystyle \int_{0}^{\pi} d\theta_1} = \frac{1}{\sqrt{\pi}} \frac{\displaystyle \Gamma \left ( d +\frac{1 - k}{2} \right ) }{\displaystyle \Gamma \left (1 + d - \frac{k}{2} \right ) } \underset{k \to \infty}{\to} \sqrt{\frac{2}{\pi (2 d - k)} }
\end{equation}

\subsubsection{Speed of convergence for the generic case}

By combination of the two previous results, if we only focus on the case $k \in [1,d]$ (no partial optimum) the comparative speed of the two research strategies, defined as the ratio of the length of each step divided by the computation time at each step, is:
\begin{equation}
{\rm acceleration}_{\rm single \ possibility} = \frac{\langle {\rm speed} \rangle_{PRW}}{\langle {\rm speed} \rangle_{\rm gradient}} = \frac{\displaystyle \frac{\langle {\rm length} \rangle}{\langle {\rm step} \rangle}}{\displaystyle \frac{1}{2 d}}  \underset{k \to \infty}{\to} \sqrt{\frac{2}{\pi} k}
\end{equation}

The limit does not work for $k = 1$ (single possibility assumption) since the Stirling' approximation is not valid in this case, but the overall dependance in $d$ or $k$ works.

\subsubsection{Effect of persistence}

The use of a persistence time to continue in a given direction once a descending one is chosen is based on research strategies on Euclidian lattices. It has been shown\cite{Tejedor:2012} that the mean first passage time, namely the average time to find for the first time a given target during a random walk, scales as $1/(2 l_p)$ for small $l_p$, where $l_p$ is the persistence length. This result is valid for a random walk on a finite Euclidian lattice of dimension $d$ with a typical size $X$: the $l_p$ that minimize the mean first passage time, $l_p^*$ scales then as a submultiple of $X$. In our case, the lattice is oriented and could be infinite on some dimensions if the optimum is at $- \infty$ or $+ \infty$.

The previous result is however a minimum: in our case the random walk is ``oriented'' to the optimum with unidirectional links between each node. Persistence can limit the number of computation of $f$ to 1, with no influence on the efficiency of the direction chosen.

For the locally quadratic case, persistence can allow to choose a descending direction on the first try, switching the $\langle {\rm step} \rangle$ from $2 (1 - 1/d)$ to $1$. This lead to a speed of convergence of at most $2 \sqrt{2/\pi d}$ for large $d$.

\section{Numerical computation for quadratic functions}

To confirm the previous asymptotic scaling behavior, numerical simulations have been performed on quadratic functions in dimension $d$. For each function, 100 starting points are chosen. For each starting point, a gradient descent method is performed and the number of steps is multiplied by $d$ to obtain the number of times that $f$ is computed. In the same time, 10 000 iterations of the persistent random walkers method are performed and the average number of steps is computed. 

\begin{figure}[htb!]
\centering \includegraphics[width = 0.8\linewidth,clip]{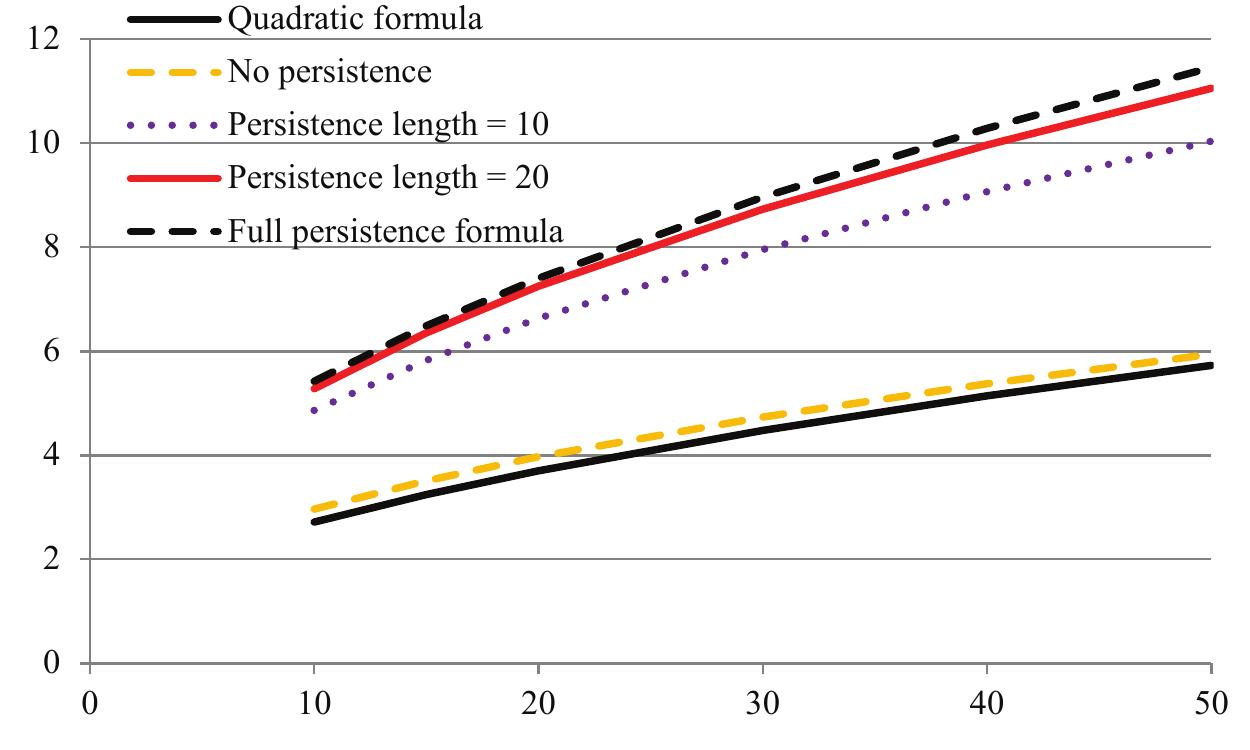} 
\caption{(color online) Acceleration factor for a persistent random walker method {\it versus} a gradient descent method as a function of the dimension $d$ of the space. The random walker method is performed with no persistence (yellow dashed line), a persistence length of 10 (violet dotted line) and of 20 (red continuous line). The theoretical lines are the quadratic formula with no persistence (black continuous line) and with a full persistence effect (black dashed line).}
\label{fig:simu}
\end{figure}

The acceleration factor with the persistent random walkers {\it versus} the gradient descent is then compared to the $\sqrt{2/\pi d} ( 1 + 3/(4 d))$ (without the persistence effect) and to $2\sqrt{2/\pi d}( 1 + 3/(4 d))$ (with the maximal persistence effect). As shown in Figure \ref{fig:simu}, the overall behaviour in $\sqrt{d}$ is well described, and the lower and upper theoretical limits gives a good approximation of the computed values.

\section{Conclusion}

The persistent random walker method introduced here allow to accelerate an optimum research by a factor $\sqrt{d}$ where $d$ is the space dimension of the function $f$. This method can be applied advantageously when the computation of $f$ is time-consuming and the space dimension $d$ is large. Applications range then from theoretical chemistry\cite{Cramer:2004} to machine learning or deep learning\cite{Bottou:1998}.

\bibliographystyle{plain}

\bibliography{../../biblio}
\end{document}